# Identification des lois de comportement des tôles à partir des essais de gonflement hydraulique

Ali KHALFALLAH[a], Seif-Allah ZOUEGUI[a]

[a] Laboratoire de Génie Mécanique, Ecole Nationale d'Ingénieurs de Monastir, Av. Ibn El-Jazzar, 5019-Monastir, Tunisie, khalfallah_a@yahoo.fr

**Résumé –** Le papier présente une nouvelle méthode pour l'identification des lois de comportement des tôles minces anisotropes à partir des essais de gonflement hydraulique en utilisant les essais à matrices circulaires et elliptiques. Cette méthode est basée sur un modèle analytique utilisant la théorie des membranes et des mesures expérimentales correspondants à la hauteur au pôle en fonction de la pression hydraulique appliquée. Quatre essais de gonflement hydraulique sont utilisés pour l'identification de la courbe d'écrouissage et les coefficients d'anisotropie du critère de Hill48. Une étude de sensibilité des paramètres matériels sur les réponses des essais de gonflement hydraulique (pression, hauteur et épaisseur au pole) pour les quatre essais est menée par simulation numérique par éléments finis. Cette méthodologie d'identification est appliquée à un acier doux DC04 destiné à l'emboutissage. Les résultats de l'identification obtenus sont utilisés pour simuler l'essai de traction plane. Il est démontré que les résultats trouvés sont en bon accord avec les réponses expérimentales.

**Mots clés :** Identification /lois de comportement / gonflement hydraulique/ matrice elliptique/ Tôles minces anisotropes.

**Abstract –** The paper describes a new method for the identification of the flow stress curves of anisotropic sheet metals using a hydraulic bulge tests through circular and elliptical dies. This method is based on analytical model using the membrane equilibrium equation and experimental data involving the measurement of only polar deflection and applied hydraulic pressure. Four hydraulic bulge tests are used for the identification of flow stress parameters and anisotropy coefficients of Hill48 yield criterion. A sensitivity analysis of material parameters is carried out by FEA of hydraulic bulge tests. This identification procedure is applied on low carbon steel DC04 used for sheet metal forming. The obtained results are used for numerical simulation of plane tensile test to validation the proposed method. It is shown a good agreement between predicted and experimental results.

## 1 Introduction

L'essai de gonflement hydraulique est un essai biaxial qui peut être utilisé comme une alternative à l'essai de traction simple. La déformation plastique homogène en traction simple est limitée par l'apparition de la striction qui se manifeste relativement « tôt » par rapport à celle observée dans l'essai de gonflage. L'extrapolation de la courbe d'écrouissage obtenue à partir de l'essai de traction simple pour des larges déformations ne permet pas de prédire correctement la formabilité et le retour élastique dans les procédés de mise en forme des tôles. En effet, l'essai de gonflement hydraulique a suscité l'attention des chercheurs. Des analyses de l'essai du gonflement hydraulique à travers des matrices circulaire et elliptiques font l'objet de travaux publiés dans la littérature [1,2]. Mais peu de recherches sont dédiées à l'identification des paramètres de comportement des tôles anisotropes à partir des essais de gonflement hydraulique. Dans ce contexte, ce papier présente une méthode combinant un modèle analytique et des simulations numériques de l'essai de gonflage, pour identifier la loi d'écrouissage et les coefficients d'anisotropie. Les essais de gonflement hydraulique à travers des matrices elliptiques sont effectués en tournant dans chaque essai le grand axe de l'ellipse par rapport à la direction de laminage pour capturer l'anisotropie du matériau.

## 2 Modèle analytique

L'essai de gonflement hydraulique consiste à appliquer une pression par un fluide pour déformer un flan retenu à ses extrémités à travers une cavité débouchante. Figure 1 représente la géométrie du flan gonflé à travers une matrice elliptique de grand axe a et de petit axe b. L'analyse de l'essai pour une matrice circulaire est déduite en égalisant les deux axes de l'ellipse (a=b).

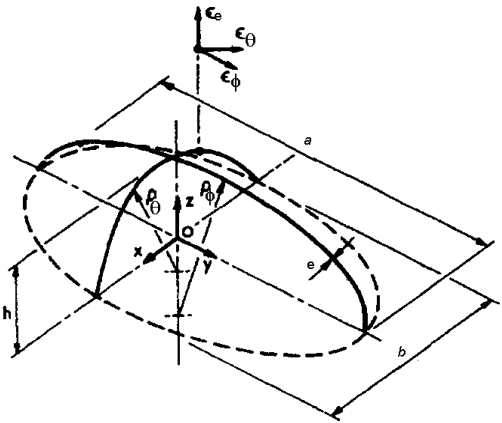

**Figure 1.** Représentation de la géométrie du flan hydroformé à travers une matrice elliptique.

L'analyse de l'essai de gonflement est basée sur l'équilibre d'un élément infinitésimal au pôle. Nous considérons le cas où le grand axe de l'ellipse coïncide avec la direction du laminage de la tôle. Les déformations plastiques suivant les trois directions principales sont obtenues à partir de l'épaisseur.

$$\varepsilon_e = \ln\frac{e}{e_0},\ \varepsilon_\theta = -\frac{\beta}{1+\beta}\varepsilon_e\ \text{et}\ \varepsilon_\phi = -\frac{1}{1+\beta}\varepsilon_e \quad (1)$$

Où $e$, $e_0$ représentent les épaisseurs instantanée et initiale. $\beta$ désigne le rapport des déformations.

Les contraintes biaxiales selon les deux directions ($\phi$ : grand axe et $\theta$ : petit axe de l'ellipse). La théorie des membres permet de déterminer les contraintes biaxiales suivant les deux directions du grand axe a et petit axe b de la matrice elliptique.

$$\sigma_\theta = \frac{P}{e}\left(\frac{1}{\rho_\theta} + \frac{\alpha}{\rho_\phi}\right)^{-1},$$

$$\sigma_\phi = \frac{\alpha P}{e}\left(\frac{1}{\rho_\theta} + \frac{\alpha}{\rho_\phi}\right)^{-1},\ \frac{\sigma_\phi}{\sigma_\theta} = \alpha \quad (2)$$

Où $\rho_\phi, \rho_\theta$ désignent les rayons de courbures selon le grand et le petit axe, P est la pression appliquée. $\alpha$ représente le rapport des contraintes. Pour calculer ces contraintes, nous devons mesurer les rayons de courbes et l'épaisseur au pole.

Les rayons de courbures selon le grand axe a et le petit axe b sont déterminés d'après Panknin [3] en tenant compte du rayon du congé de la matrice $r_f$, de la hauteur au pole $h$ et des longueurs des axes a et b.

$$\rho_\phi = \frac{(a+r_f)^2 + h^2 - 2r_f h}{2h},\ \rho_\theta = \frac{(b+r_f)^2 + h^2 - 2r_f h}{2h} \quad (3)$$

Pour calculer l'épaisseur, plusieurs relations sont proposées dans la littérature. Nous choisissons de déterminer cette épaisseur à par de la simulation par la méthode des éléments finis de l'essai de gonflement hydraulique. La contrainte équivalente et déformation équivalente sont obtenues à partir du critère orthotrope de Hill48. La figure 2 montre l'acheminement des calculs pour aboutir à la contrainte et déformation équivalente.

$$\overline{\sigma} = c\left(\frac{r_0\sigma_\theta^2 + r_{90}\sigma_\phi^2 + r_0 r_{90}(\sigma_\theta - \sigma_\phi)^2}{r_0 r_{90}}\right)^{1/2}$$

$$c = \frac{3}{2(1+(1/r_0)+(1/r_{90}))} \quad (4)$$

$$\overline{\varepsilon} = \frac{c1}{c2}\left(\begin{array}{l}(1/r_{90})((\varepsilon_\theta/r_0)-\varepsilon_e)^2 + (1/r_0)(-\varepsilon_\phi/(1+r_0)+\varepsilon_e)^2 \\ + ((\varepsilon_\phi/r_{90})-(\varepsilon_\theta/r_0))^2\end{array}\right)^{1/2}$$

$$c_1 = (2/3)(1+1/r_0 +1/r_{90})$$
$$c_2 = ((1/r_0 r_{90}) +1/r_0 +1/r_{90})) \quad (5).$$

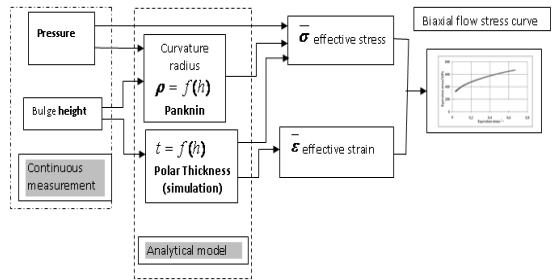

**Figure 2.** Méthodologie pour calculer la contrainte et la déformation équivalente.

Pour déterminer les contraintes et déformations équivalente pour les autres essais, dont le grand axe est orienté à direction 90° et 45° par rapport à la direction du laminage, il suffit de tenir compte de la rotation du repère d'orthotropie par rapport au repère principale des contraintes et déformation définie par les directions selon le grand et petit axe de l'ellipse.

## 3 Etude numérique de sensibilité

Une étude de sensibilité est menée pour investiguer l'influence des paramètres matériels tels que le coefficient d'écrouissage, $n$ les coefficients d'anisotropie $r_0, r_{45}$ et $r_{90}$ sur la pression et l'épaisseur en fonction de la hauteur au pole. Ces réponses sont obtenues par simulation numérique par la méthode des éléments finis des quatre essais de gonflement hydraulique, circulaire et elliptique. Cette étude sensibilité permet de justifier la pertinence de ces essais pour l'identification de la loi d'écrouissage et l'anisotropie à partir des essais de gonflement hydraulique. Les résultats obtenus montrent que le coefficient d'écrouissage influe sur l'évolution de la pression en fonction de la hauteur au pole et sur l'épaisseur et ceci pour les quatre essais d'hydroformage. Cependant, les coefficients d'anisotropie n'influent pas sur l'épaisseur au pole quelque soit l'essai, mais ils influent sur la pression en fonction de la hauteur au pole. En effet, l'épaisseur utilisée pour calculer les contraintes et déformations équivalentes est générée par simulation numérique en utilisant les propriétés de l'essai de traction simple.

## 4 Essais expérimentaux

Des essais de gonflement hydraulique avec matrice circulaire et elliptique ont été réalisés [4] pour identifier la loi d'écrouissage et l'anisotropie d'un acier doux DC04, d'épaisseur 1mm. La figure 3 montre les quatre flans hydroformés jusqu'à la rupture et correspondant aux quatre essais numéroté de 0 jusqu'à 3. L'essai 0 correspond à l'essai avec matrice circulaire qui à un diamètre de 91 mm et un rayon du congé $r_f$ = 6mm. Les essais 1,2 et 3 correspondent aux essais avec la matrice elliptique qui a un grand axe de longueur a= 110 mm, un petit axe de longueur b= 74 m et un rayon du congé $r_f$ = 6mm. En outre, des essais de traction simple sont réalisés sur ce matériau.

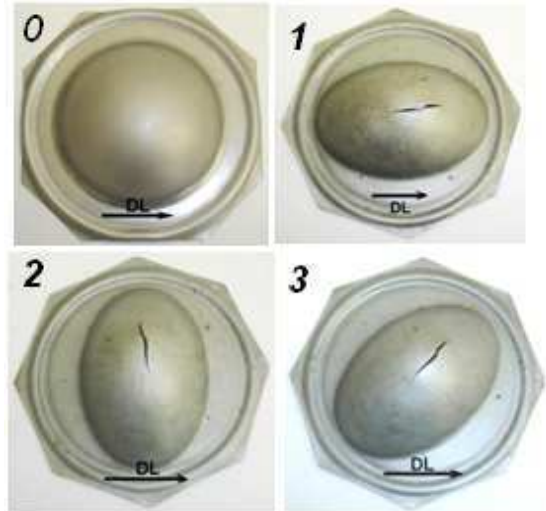

**Figure 3.** Les flans hydroformés à travers une matrice circulaire

## 5 Méthode d'identification des paramètres de comportement

La méthode d'identification de paramètres de comportement de l'acier DC04 est découplée en deux étapes. La première étape consiste à déterminer les coefficients d'anisotropie du critère de Hill48 et la deuxième étape permet d'identifier les paramètres de la loi d'écrouissage de Swift.

### Identification de l'anisotrope

Pour identifier les coefficients d'anisotropie du modèle du modèle de Hill48, les courbes des contraintes équivalente correspondant à chaque essai (essais 0,1,2 et 3) sont calculées (c.f. Figure 2). Le problème est ramené à un problème d'optimisation en minimisant une fonction objective définissant l'écart entre les quatre courbes de la contrainte équivalente en fonction de la déformation équivalente

$$f = \sum_{i=1}^{N}\left((\bar{\sigma}_{cir}-\bar{\sigma}_{e1})^2 +(\bar{\sigma}_{cir}-\bar{\sigma}_{e2})^2 +(\bar{\sigma}_{cir}-\bar{\sigma}_{e3})^2\right) \quad (6)$$

La méthode d'optimisation choisie pour résoudre ce problème est une méthode stochastique. Il s'agit de la méthode recuit simulé programmée sous Matlab.

La figure 4 représente les courbes contraintes équivalentes-déformations équivalentes pour les quatre essais de gonflement hydraulique, circulaire et elliptique (essais 0,1, 2 et 3) obtenues après la convergence de la procédure d'identification. On observe que les courbes correspondant aux essais de gonflement hydraulique circulaire et elliptique sont en bon accord avec la courbe d'écrouissage obtenue à partir de l'essai de la traction simple. Le tableau 1 liste les coefficients d'anisotropie identifiés et ceux mesurés par la traction simple.

**Tableau 1.** Coefficients d'anisotropie identifiés et mesurés expérimentalement

|  | Coefficient d'anisotropie | | |
|---|---|---|---|
|  | *r0* | *r45* | *r90* |
| Gonflement | 1.15 | 0.8 | 1.85 |
| Traction simple | 1.59 | 1.00 | 1.54 |

### Identification des paramètes d'écrouissage

La loi d'écrouissage de Swift est utilisée pour modéliser l'écrouissage isotrope de l'acier DC04.

$$\overline{\sigma} = K(\varepsilon_0 + \overline{\varepsilon})^n \quad (7)$$

L'identification des trois paramètres de cette loi est obtenue en effectuant un simple lissage des points de la courbe d'écrouissage correspondant à l'essai de gonflement hydraulique à travers la matrice circulaire, après la convergence de la procédure d'identification. Le tableau 2 présente les coefficients de la loi d'écrouissage de Swift obtenus à partir des essais de gonflement hydraulique et de l'essai de traction simple dans la direction 0°. Les jeux de paramètres correspondant aux deux types d'essais sont du même ordre de grandeur.

**Tableau 2.** Coefficients de la loi de Swift

|  | Loi de Swift | | |
|---|---|---|---|
|  | *K[MPa]* | $\varepsilon_0$ | *n* |
| Gonflement | 679,53 | 0,03 | 0,32 |
| Traction simple | 619,27 | 0,02 | 0,24 |

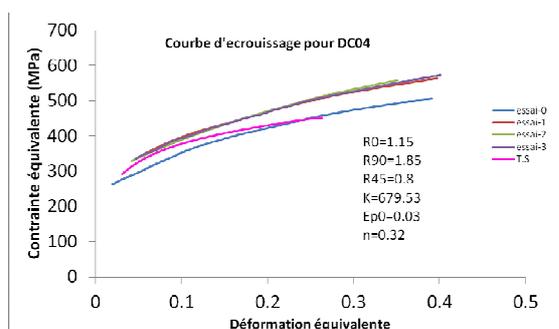

**Figure 4.** Courbe d'écrouissage identifiée

## 6 Validation de la méthode par l'essai de traction plane

Pour valider la méthode proposée pour identifier la loi d'écrouissage et l'anisotropie d'un acier doux à partir des essais de gonflement hydraulique circulaire et elliptique, l'essai de la traction plane est utilisé. Les coefficients matériels identifiés sont utilisés pour simuler l'essai de traction plane suivant trois direction 0°, 45° et 90° par rapport à la direction de laminage. La figure 5 représente les résultats prédits et les réponses expérimentales. Ces réponses prédites par simulation numérique de l'essai de la traction plane sont comparées aux résultats expérimentaux.

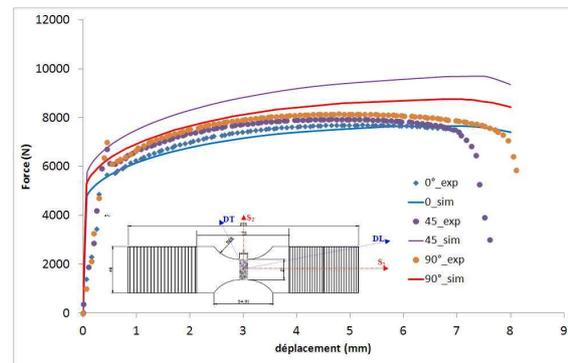

**Figure 5**. Validation par l'essai de traction plane.

## 7 Conclusion

Une méthode d'identification de la loi d'écrouissage et d'anisotropie à partir des essais de gonflement hydraulique circulaire et elliptique est proposée. Cette méthode a l'avantage d'utiliser deux grandeurs expérimentales : la pression et la hauteur au pole. L'épaisseur est simulée et les rayons de courbures sont calculés. Les résultats obtenus sont en bon accord avec l'expérimentale.

### Bibliographie